\documentclass{Interspeech2024}
\usepackage{kotex} 

\interspeechcameraready

\title{Period Singer: Integrating Periodic and Aperiodic Variational Autoencoders for Natural-Sounding End-to-End Singing Voice Synthesis}

\name[affiliation={}]{Taewoo}{Kim}
\name[affiliation={}]{Choongsang}{Cho}
\name[affiliation={}]{Young Han}{Lee}

\address{
    Korea Electronics Technology Institute, South Korea
}
\email{\{kimtaewoo, ideafisher, yhlee\}@keti.re.kr}


\begin{document}

\maketitle

\begin{abstract}

In this paper, we present Period Singer, a novel end-to-end singing voice synthesis (SVS) model that utilizes variational inference for periodic and aperiodic components, aimed at producing natural-sounding waveforms.
Recent end-to-end SVS models have demonstrated the capability of synthesizing high-fidelity singing voices.
However, owing to deterministic pitch conditioning, they do not fully address the one-to-many problem.
To address this problem, we present the Period Singer architecture, which integrates variational autoencoders for the periodic and aperiodic components.
Additionally, our methodology eliminates the dependency on an external aligner by estimating the phoneme alignment through a monotonic alignment search within note boundaries.
Our empirical evaluations show that Period Singer outperforms existing end-to-end SVS models on Mandarin and Korean datasets.
The efficacy of the proposed method was further corroborated by ablation studies.

\end{abstract}

\section{Introduction}
Singing voice synthesis (SVS) aims to generate natural singing voices similar to those of humans based on music scores that include lyrics, note pitch, and note duration.
In text-to-speech (TTS) tasks, there is significant uncertainty owing to the wide variation encompassing various emotions and intonations, as it depends solely on the text.
By contrast, SVS tends to have relatively less uncertainty than TTS because it is guided by the note pitch and duration information of the music scores.
However, modeling narrow variations based on the rich expression of the singing voice is crucial but challenging for achieving natural singing voices.
Early deep learning-based SVS models~\cite{xiaoicesing, litesing} adopted a two-stage approach using a parametric vocoder. 
For instance, XiaoiceSing~\cite{xiaoicesing} successfully generated natural singing voices by leveraging the Fastspeech~\cite{fastspeech} architecture and WORLD vocoder~\cite{world}.
Subsequently, neural vocoder-based SVS models~\cite{nsinger,amtlsvs,wesinger2,xiaoicesing2,diffsinger} were proposed because of the performance ceiling of parametric vocoders, enabling the generation of high-quality singing voices.
However, the two-stage model can still be improved to address the mismatch in acoustic features between the training and inference stages.

To address this discrepancy, the end-to-end VISinger series~\cite{visinger,visinger2} has been introduced.
VISinger~\cite{visinger} was the first end-to-end singing voice synthesis model based on VITS~\cite{vits}.
Specifically, they introduced the frame prior network to effectively model narrow variations within a phoneme in singing voices.
The improved version, VISinger2~\cite{visinger2}, could successfully alleviate pitch glitch issues while achieving high fidelity at a 44.1 kHz sampling rate.
However, they did not fully solve the one-to-many problem because the models were conditioned on deterministic pitches
Moreover, it requires ground truth phoneme duration information for training because it removes the monotonic alignment search to adopt the frame prior network.

In this study, we propose a novel end-to-end SVS model called Period Singer, which enables variational pitch modeling. 
To address the one-to-many problem, we integrate conditional variational autoencoders (CVAEs) for both the periodic and aperiodic components in the proposed model.
Additionally, we apply the smoothed pitch augmentation method to ensure that the latent variables capture both wide and narrow pitch variations.
During training, we utilized a normalizing flow to transform the aperiodic posterior distribution into a linguistic representation of a simpler distribution. 
This allows for the monotonic alignment search while adopting frame prior networks. 
To enhance the monotonic alignment search, we utilized note boundaries to prevent mismatches that lead to meaningless alignment between the singing voice and the music score.
In the experiments, we evaluated our model on the public Mandarin dataset, Opencpop~\cite{opencpop} and the internal Korean dataset, demonstrating superior performance compared to existing end-to-end models. 
Moreover, we validated the effectiveness of the proposed approach through ablation studies.

\section{Proposed methods}
\begin{figure*}[t]
  \centering
  \includegraphics[width=1.0\linewidth]{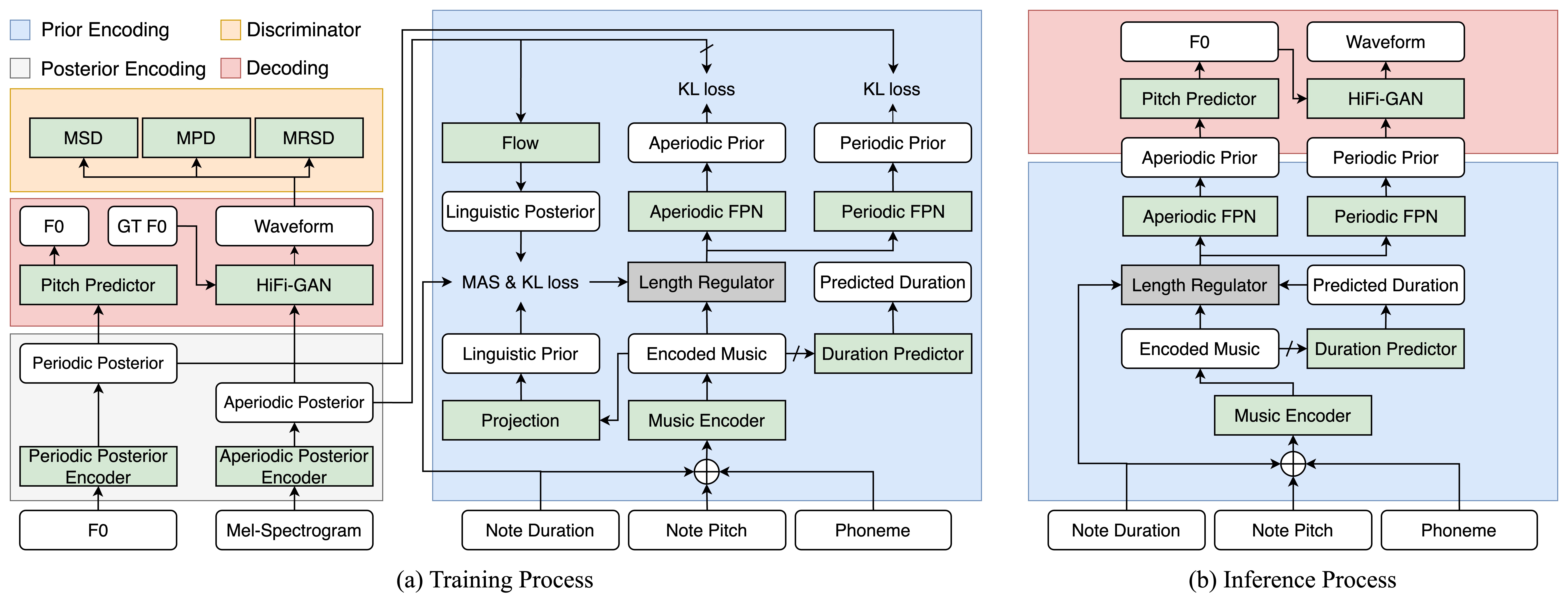}
  \caption{Overall architecture of Period Singer.}
  \label{fig:architecture}
  \vspace{-0.3cm}
\end{figure*}

The proposed model architecture builds on the foundation of VITS~\cite{vits} by integrating two CVAEs.
The proposed model comprises individual posterior encoders and decoders for both periodic and aperiodic components, a prior encoder, and discriminators.
The posterior encoders extract latent variables for both the periodic and aperiodic components while being constrained by the prior encoder.
In the decoding stage, periodic latent variables are utilized by the periodic decoder to predict fundamental frequency ($F_0$) sequences.
The predicted $F_0$ sequences along with the aperiodic latent variables are conditioned on the aperiodic decoder to generate the waveform.
Subsequently, the generated waveform was adversarially trained using discriminators
The overall architecture of Period Singer is shown in Figure \ref{fig:architecture}, which illustrates both the training and inference processes.

\subsection{Posterior encoding}
We apply CVAEs separately to model the periodic and aperiodic components of singing voices, using two posterior encoders.
First, the posterior encoder for the aperiodic component uses the mel-spectrogram processed from the waveform as the input, whereas the other posterior encoder uses the $F_0$ sequence as the input.
Each posterior encoder is structured similarly to VITS, comprising non-causal WaveNet residual blocks~\cite{wavenet} and a linear projection layer. It generates parameters for the mean and variance of each posterior distribution.
They are then sampled using a reparameterization trick and passed to the decoder.

\subsection{Prior encoding}
The prior encoder derives the latent variables of the linguistic, periodic, and aperiodic prior distributions from the phoneme alignment $A$ and the music score condition $c$, which includes the lyrics, note pitch, and note duration.
For CVAEs, the periodic prior distribution $p(z_{p}|c,A;\theta_p)$ and the linguistic prior distribution $p(z_{l}|c,A,\theta_l)$ are regularized.
In addition, the aperiodic prior distribution $p(z_{a}|c,A;\theta_a)$ is optimized to match the detached aperiodic posterior distribution.

As shown in Figure \ref{fig:architecture}(a), the prior encoder structure comprises a music encoder, a duration predictor, a length regulator, frame prior networks (FPNs) for periodic and aperiodic components, and a flow module.
Initially, the aperiodic posterior distribution $q(z_{a}|x_{mel};\phi_a)$ is transformed into a simplified distribution $q(z_{l}|x_{mel};\phi_l)$ for the linguistic component using the flow module.
Then, the monotonic alignment search (MAS)~\cite{vits, glowtts} is employed between the prior distribution $p(z_{l}|c,A,\theta_l)$ extracted from the music encoder and the projection layer, and the posterior distribution $q(z_{l}|x_{mel};\phi_l)$ to estimate the phoneme alignment $A$.
In the training process, the phoneme duration $d$, derived from the predicted phoneme alignment $A$, is utilized to expand the encoded music representation $h$ and linguistic prior distribution $p(z_{l}|c,A,\theta_l)$ via the length regulator. 
Conversely, during inference, the predicted phoneme duration $\hat{d}$ obtained from the duration predictor is used.
Subsequently, the expanded encoded music representation $h$ is used to extract prior distributions for periodic and aperiodic frames from the FPNs. 
The linguistic prior distribution $p(z_{l}|c,A,\theta_l)$ is also expanded for aperiodic CVAE training.

To regularize the periodic and aperiodic CVAEs, the Kullback-Leibler (KL) divergence losses $L_{kl,p}$ and $L_{kl,a}$ are calculated as follows:
\begin{align}
L_{kl,p} &= D_{kl}(q(z_p|F_0;\phi_{p})|| p(z_p|c,A;\theta_{p})), \\
L_{kl,a} &= D_{kl}(q(z_l|x_{mel};\phi_{l})|| p(z_l|c,A;\theta_{l})) \nonumber \\
&\quad + \lambda_{l} D_{kl}(q(\bar{z}_a|x_{mel};\phi_{a})|| p(z_a|c,A;\theta_{a})), 
\end{align} 
where $x_{mel}$, $\lambda_{l}$, and $\bar{z}_a$ represent the input mel-spectrogram, the weighting factor and the detached aperiodic posterior latent variables, respectively.
In addition, for the duration loss $L_{dur}$, we employ an L2 loss and apply the stop-gradient, as indicated by the short line in Figure \ref{fig:architecture}(a).

\subsubsection{Monotonic alignment search with note boundary}
In the TTS task, the monotonic alignment search (MAS) has been successfully adopted in \cite{vits, glowtts} to estimate phoneme alignment without an external aligner. 
However, the alignment estimation in singing tasks is challenging because of the presence of breath sounds and diverse variations in pitch. 
This challenge arises from the end-to-end nature of the model, which incorporates reconstruction loss for segmented features, potentially leading to meaningless mappings between the singing voices and music scores.
To address this issue, we perform MAS within note boundary information, representing the start and end frames of each note, and integrate it to accurately estimate the alignment.
In particular, when there is only one phoneme per note, it is directly enforced in the alignment without the MAS.

\subsubsection{Length regulator during inference}
The duration predicted by the duration predictor does not guarantee perfect alignment with the length of the singing voice in rhythm. 
Hence, rhythm adjustment based on note duration is essential. 
This involves calculating the ratio of the phoneme durations within a single note and scaling them to match the note duration. 
This process enables the inference of an accurate singing voice synchronized with the rhythm.

\subsection{Decoding}
Recently, singing voice vocoders~\cite{sifigan, singgan, periodhifigan} have been proposed to alleviate the glitch caused by long vowels in singing voices, which are commonly conditioned on sine waveforms.
Therefore, we employ the pitch-conditioned HiFi-GAN~\cite{periodhifigan, periodvits} as the aperiodic decoder.
The pitch predictor reconstructs $F_0$ from periodic latent variables, which are then conditioned by the aperiodic decoder. 
During training, the ground truth $F_0$ is conditioned.
Consequently, we use the mel-spectrogram loss $L_{mel}$ and the pitch loss $L_{pitch}$, both calculated using the L1 loss.

In the SVS task, because the main pitch is determined by the input notes, modeling narrow variations in pitch, such as bending and vibrato, is important.
In this study, we propose a pitch augmentation method using smoothed $F_0$ to guide the periodic CVAE to learn narrow variations in pitch.
Through the median filter, short-term bending and vibrato are removed, resulting in a smoothed $F_0$ contour.
To guide the mean of the periodic latent variables for learning the smoothed $F_0$, we concatenate the mean with the input batch of the aperiodic decoder and the smoothed $F_0$ with the target batch, respectively.
Therefore, the pitch loss is modified as follows:
\begin{align}
L_{pitch} &= ||\log{F_0} - \log{\hat{F_0}} ||_1 \nonumber \\
&\quad + \lambda_{s} || \log{f_s(F_0)} - \log{\tilde{F_0}}||_1,
  \label{equation:pitch}
\end{align}
where $\lambda_{s}$, $f_s(\cdot)$, $\hat{F_0}$, and $\tilde{F_0}$ represent a weighting factor for the smoothed $F_0$, the median filter, the predicted $F_0$, and the predicted smoothed $F_0$, respectively.

\subsection{Discriminator}
To generate high-fidelity singing voice, it is crucial to address high-frequency aliasing and artifacts.
VISinger2~\cite{visinger2} addresses this issue by combining the multi-period discriminator (MPD) and multi-scale discriminator (MSD) from HiFi-GAN~\cite{hifigan} with the multi-resolution spectrogram discriminator (MRSD) of UnivNet~\cite{univnet}. 
We also employ the same discriminator configuration, with the loss for adversarial training defined as follows:
\begin{align}
  L_{dis} &= \mathbb{E}_{(y,z_a,F_0)} \left[(D(y)-1)^2 + (D(G(z_a,F_0)))^2 \right],  \\ 
  L_{adv} &= \mathbb{E}_{(z_a,F_0)} \left[(D(G(z_a,F_0))-1)^2 \right],  \\ 
  L_{fm} &= \mathbb{E}_{(y,z_a,F_0)}\left[ \sum_{l=1}^{L}\frac{1}{N_l}||D^l(y)-D^l(G(z_a,F_0))||_1 \right],
  \label{equation:gan}
\end{align}
where $z_a$ denotes the aperiodic posterior latent variable.
$L$, $D_l$, and $N_l$ represent the number of layers, the $l$-th layer, and the number of features in that layer of the discriminator, respectively.

\subsection{Final loss}
The final loss function for optimizing the proposed model, considering the periodic and aperiodic CVAEs, as well as GAN training objectives, is as follows:
\begin{align}
  L_{gen} &= L_{adv} + \lambda_{fm} L_{fm} + \lambda_{mel} L_{mel} + \lambda_{pitch} L_{pitch} \nonumber\\ 
  &\quad +  \lambda_{a} L_{kl,a} + \lambda_{p} L_{kl,p} + \lambda_{dur} L_{dur}
  \label{equation:final}
\end{align}
where the $\lambda$ coefficient serves as a weighting factor for each loss.

\section{Experiments}
\subsection{Datasets}
In the experiments, we evaluated the performance of Period Singer separately for two languages: Mandarin and Korean.
\begin{itemize}
\item \textbf{Mandarin}: We utilized the publicly available singing corpus, Opencpop~\cite{opencpop}, which comprises recordings from a single female singer totaling 100 songs and includes detailed annotations such as lyrics, notes, phoneme alignments, silence, aspirates, and slurs.
We used predefined training and testing sets from Opencpop, consisting of 95 songs (3,550 segments) and 5 songs (206 segments), respectively.
\end{itemize}

\begin{itemize}
\item \textbf{Korean}: We selected 100 Korean pop songs and collected singing datasets recorded by a female artist. 
The dataset was divided into 95 songs (2,461 segments) for training and 5 songs (127 segments) for testing, with each segment ranging from 3 to 12 seconds.
It included lyrics and MIDI notes manually annotated by human annotators.
\end{itemize}

All audio files across both datasets were sampled at 44.1 kHz with 16-bit quantization.
In the preprocessing stage, an 80-mel-spectrogram was extracted with settings of 2048 for the FFT size, 2048 for the window size, and 512 for the hop size. 
The $F_0$ extraction step was performed using the harvest~\cite{harvest} algorithm from the WORLD toolkit~\cite{world}. 
For pitch augmentation, we extracted the smoothed $F_0$ using a median filter with a kernel size of 13.

\subsection{Model configurations}
The proposed Period Singer architecture was developed based on VITS~\cite{vits}, incorporating additional components such as the periodic and aperiodic FPNs, the periodic posterior encoder, and the pitch predictor. 
The FPNs for periodic and aperiodic components consisted of 6 Conformer blocks~\cite{conformer} and a projection layer.
Each Conformer block was set with 2 heads, a hidden size of 192, and a kernel size of 31. 
Posterior encoders were composed of 8 Wavenet residual blocks~\cite{wavenet} with a kernel size of 5. 
The mean and variance of all the distributions were 192-dimensions. 
The structure of the pitch predictor mirrored that of the posterior encoder, comprising 4 Wavenet residual blocks and a projection layer, with the output size set to 1. 
Our decoder module followed the pitch-conditioned HiFiGAN architecture of~\cite{periodvits}, with 256 hidden dimensions. 
To extend the proposed model to 44.1 kHz, the upsampling stride and kernel sizes of the HiFi-GAN module~\cite{hifigan}, originally set at 22.05 kHz, were adjusted to (8,8,4,2) and (16,16,8,4) respectively. 
The discriminator model configuration was the same as that of VISinger2, while the remaining hyperparameters were consistent with those of VITS.
For the weights in the final loss, we set $\lambda_{l}$, $\lambda_{s}$, $\lambda_{fm}$, $\lambda_{mel}$, $\lambda_{pitch}$, $\lambda_{a}$, $\lambda_{p}$ and $\lambda_{dur}$ to 1, 1, 2, 45, 10, 1, 1, and 1, respectively.

During training, we used the AdamW optimizer~\cite{adamw} with $\beta_1=0.8$, $\beta_2=0.99$ and weight decay $\lambda=0.01$.
The learning rate decayed exponentially by a factor of $0.999^{1/8}$ for each epoch with an initial learning rate of $0.0002$.
The proposed model was trained for 400k iterations with a batch size of 8.

\subsection{Baseline systems}
For comparison, we trained the following baseline systems:

\begin{itemize}
\item \textbf{VITS}: Original VITS trained with music score input. We utilized the official implementation\footnote{\url{https://github.com/jaywalnut310/vits}} with slight modifications for music score input.
\item \textbf{VISinger}: The end-to-end SVS system based on VITS. It predicts deterministic F0 and conditions it on the frame prior network. We utilized the unofficial implementation\footnote{\url{https://github.com/jerryuhoo/VISinger}}. 
\item \textbf{VISinger2}: The improved system of VISinger that predicts the deterministic F0 and mel-spectrogram, and conditions it on the frame prior network and the decoder. We utilized the official implementation\footnote{\url{https://github.com/zhangyongmao/VISinger2}}, setting the number of the FFT block to 6 instead of the configured 4 as in \cite{visinger2}.
\item \textbf{Period Singer (DPP)}: The Period Singer model with deterministic pitch prediction (DPP). The periodic frame prior network directly predicts $F_0$ and conditions it on the decoder.
\end{itemize}

All baseline systems were extended to a sampling rate of 44.1 kHz and trained with the same discriminator and configuration as the Period Singer.
To train VISinger and VISinger2 on the Korean dataset, we obtained phoneme durations using the proposed alignment estimation, which utilized note boundary information from a pre-trained VITS.

\subsection{Results}

\begin{table}[t]
  \centering
  \caption{The MOS test results with 95\% confidence intervals}
  \label{tab:mos}
  \begin{tabular}{@{\hspace{0.2cm}}p{3.5cm}cc@{\hspace{0.2cm}}}
    \toprule
    Model  & Mandarin & Korean \\ \midrule
    VITS & $3.67 \pm 0.08$ & $3.78 \pm 0.07$ \\
    VISinger & $3.71 \pm 0.08$ & $3.88 \pm 0.07$ \\
    VISinger2 & $3.91 \pm 0.07$ & $4.45 \pm 0.06$ \\
    Period Singer (DPP) & $3.90 \pm 0.07$ & $4.56 \pm 0.05$ \\
    Period Singer& $\textbf{3.97}$ $\pm$ $\textbf{0.07}$ & $\textbf{4.61}$ $\pm$ $\textbf{0.05}$ \\ \midrule
    Recording & $4.24 \pm 0.07$ & $4.65 \pm 0.05$ \\ \bottomrule
  \end{tabular}
\end{table}

\begin{table}[t]
  \centering
  \caption{The CMOS test results for ablation study}
  \label{tab:ablation}
  \begin{tabular}{@{\hspace{0.2cm}}p{3.5cm} @{\hspace{0.2cm}} c@{}}
    \toprule
    \multicolumn{1}{@{\hspace{0.2cm}}l}{Model} & \multicolumn{1}{c}{CMOS} \\ \midrule
    Period Singer & $0$ \\
    -- Normalizing Flow & $-0.21$ \\ 
    -- Note boundary for MAS & $-1.13$ \\
    \bottomrule
  \end{tabular}
  \vspace{-0.3cm} 
\end{table}

We conducted a mean opinion score (MOS) test to assess the overall performance, encompassing the sound quality and naturalness, of our Period Singer and baseline systems.
For the evaluation in Mandarin, 20 testers engaged via Amazon Mechanical Turk, whereas 19 Korean testers participated in the Korean assessment.
We provided 20 samples per system, and each tester rated 120 samples across the 6 systems, including the recordings.
All testers were asked to evaluate the singing voice on a scale of 1 to 5 with intervals of 0.5.

Table \ref{tab:mos} summarizes the MOS results.
For both datasets, the scores for Period Singer are 3.97 and 4.61, respectively, indicating superior performance compared to all other systems.
In particular, for the Korean dataset, Period Singer obtains scores that reached the confidence intervals of the recordings, demonstrating the sound quality and naturalness close to human recordings.
Period Singer (DPP) shows slightly lower performance compared to VISinger2 by 0.01 for the Mandarin dataset.
On the other hand, it exhibits better performance for the Korean dataset.
These outcomes are acceptable given the relatively lower diversity of singing voices in the Opencpop dataset compared to the Korean dataset.
Moreover, these results are remarkable considering that the Period Singer models were trained without ground truth phoneme duration.

\subsection{Ablation study}

To verify the effectiveness of the method proposed in Period Singer, we conducted ablation studies.
First, we compared the naturalness with and without the normalizing flow module and note boundary for the MAS. 
To this end, we prepared 20 samples for each system and requested an additional comparative MOS (CMOS) from the 19 Korean testers.
They rated other systems, compared to Period Singer, on a scale of $-3$ to $3$, with intervals of $1$.

As shown in Table \ref{tab:ablation}, removing the normalizing flow module results in a decrease of 0.21 points in performance. 
This suggests that a simpler linguistic distribution, rather than performing MAS with an aperiodic distribution, is more effective for predicting an accurate alignment. 
Furthermore, when using the original MAS without note boundaries, the performance is 1.13 points lower compared to the proposed method. 
This indicates that constraining MAS with note boundaries leads to a stable alignment prediction, avoiding meaningless mappings between the singing voice and the music score. 
Some audio samples are available online\footnote{\url{https://rlataewoo.github.io/periodsinger}}. 

\begin{figure}[t]
  \centering
  \includegraphics[width=\linewidth]{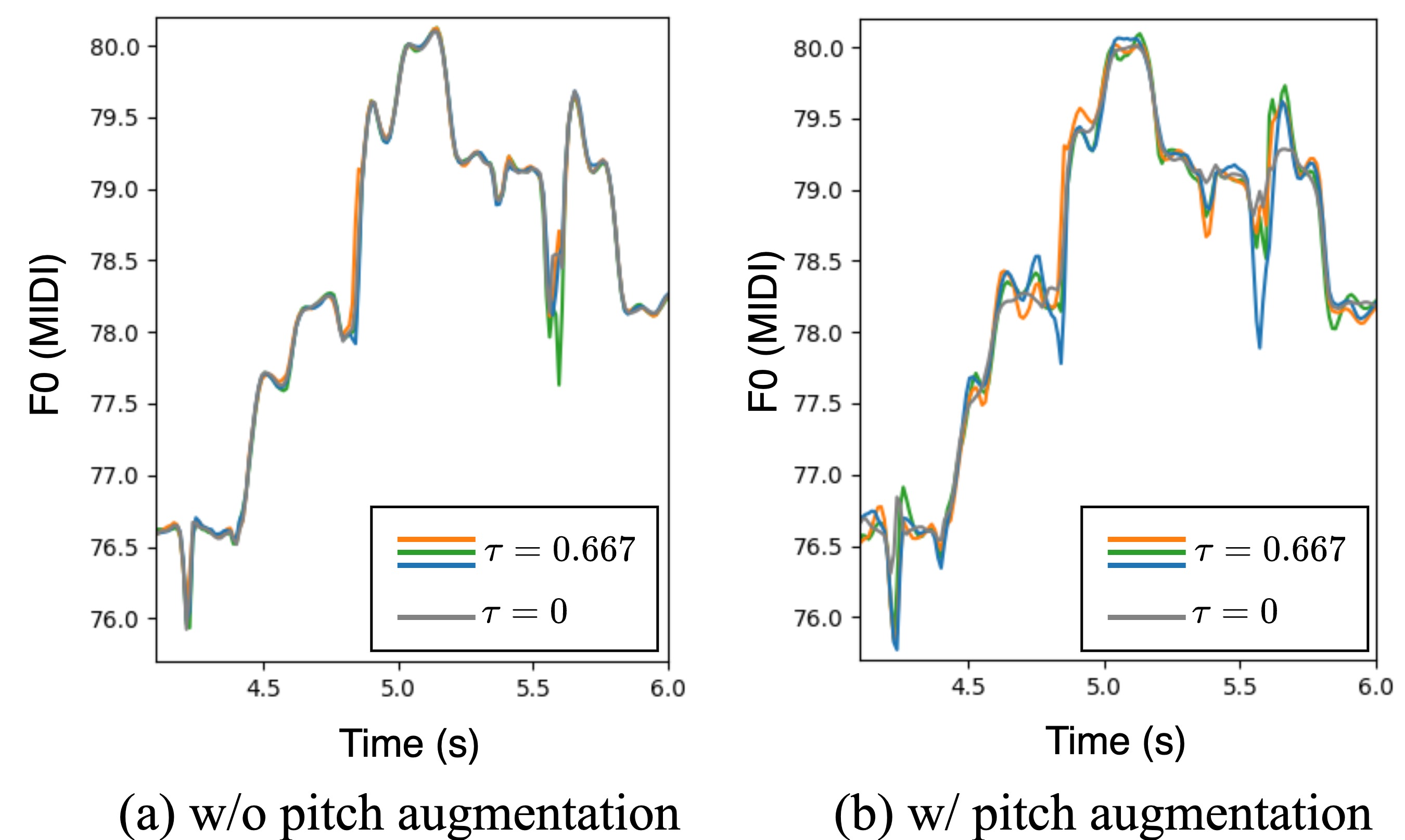}
  \caption{Variational pitch inferernce of Periodic Singer with and without pitch augmentation.}
  \label{fig:f0}
\end{figure}

To confirm the impact of variational pitch inference on pitch augmentation, we compared the results obtained with and without the proposed pitch augmentation. 
We visualized the results in Figure \ref{fig:f0} by setting the temperature $\tau$, which served as the scale factor for variance, to 0.667 and 0, respectively, and conducting the inference four times and once for each setting.
Figure \ref{fig:f0}(a) illustrates that for Period Singer without pitch augmentation, the results are similar when $\tau$ is 0 and 0.667, indicating a heavily sparse variance. 
However, as shown in Figure \ref{fig:f0}(b), when $\tau$ is zero, the predictions tend to cluster around the mean pitch, whereas, at 0.667, we observed a more diverse range of pitches compared with Figure \ref{fig:f0}(a), indicating a broader distribution of variance.

\section{Conclusions}
In this study, we introduced Period Singer, a novel end-to-end SVS model designed to produce natural and high-fidelity waveforms.
By incorporating variational inference for both periodic and aperiodic components, Period Singer addresses the limitations of previous SVS models, particularly the one-to-many problem stemming from deterministic pitch conditioning.
Our architecture not only enhances the synthesis quality but also eliminates the need for an external aligner by estimating the phoneme alignment within the note boundary information.
Empirical evaluations on Mandarin and Korean datasets demonstrate that Period Singer surpasses existing end-to-end SVS models.

\section{Acknowledgements}
This work was partly supported by Institute of Information \& communications Technology Planning \& Evaluation (IITP) grant funded by the Korea government (MSIT) (No.2022-0-00963 and No.2022-0-00608).

\bibliographystyle{IEEEtran}
\bibliography{mybib}

\end{document}